\documentstyle[preprint,aps]{revtex}
\begin{document}
\draft
\preprint{IASSNS-HEP-94/42} 
\title{
Elementary Excitations of One-Dimensional $t$-$J$ Model \\ 
with Inverse-Square Exchange
}
\author{Z. N. C. Ha$^{(1)}$ and F. D. M. Haldane$^{(2)}$ }
\address{
$~^{(1)}$School of Natural Sciences, Institute for Advanced Study,
Princeton, New Jersey 08540. \\
$~^{(2)}$Department of Physics, 
Princeton University, Princeton, New Jersey 08544. \\
}
\date{June 14, 1994}
\maketitle
\begin{abstract}
We identify exact excitation content of the intermediate states
for the one-particle Green's functions,  spin-spin
and (charge) density-density correlation functions of the periodic 
one-dimensional $t$-$J$ model with inverse square exchange.
The excitations consist of neutral $S=1/2$  spinons and 
spinless (charge $-e$) holons with
semionic fractional statistics, and bosonic (charge +2$e$) ``anti-holons''
which are excitations of the holon condensate.    
Due to the supersymmetric Yangian quantum symmetry of this model,
only the excited states with {\it finite} number of elementary 
excitations contribute to the spectral functions.
We find a set of selection rules, and this allows us to map out 
the regions of non-vanishing spectral weight
in the energy-momentum space for the various correlation functions.
\end{abstract}
\narrowtext

Recently, there have been many developments  in understanding the family of
Calogero-Sutherland models (CSM) which are identified with their 
peculiar inverse-square exchange (ISE)
\cite{carlo,haldane2,kura,yangian,lattice,sun1,sun2,inozem}.
An important feature of these models is that
they belong to the same low-energy universality class as the family of
Bethe-ansatz solvable models and may provide a new fully soluble 
paradigm next to the non-interacting models \cite{haldane2}.  

The one-dimensional supersymmetric ISE $t$-$J$ model \cite{kura} 
represents a fixed point model where the elementary excitations
form an ideal gas obeying fractional statistics.  In contrast to
this model,
the NNE $t$-$J$ model \cite{tjmod,bares}, which has essentially the 
same low energy spectra spanned by the same elementary excitations,
obscures the simple low energy structure intrinsic to this 
class of models.  We rediscover the spinons, the holons and the
antiholons---the elementary excitations of the NNE 
$t$-$J$ model \cite{bares}---in the context of
the supersymmetric Yangian of the ISE model.  
Furthermore, we find that
only the states with {\it finite} number of these elementary
excitations contribute to the spectral functions of
the one-particle Green's functions ($G^{(1)}$), the charge 
density-density ($C^{(c)}$) and the spin-spin 
correlation functions ($C^{(s)}$).

First, we examine the symmetry in the ISE supersymmetric $t$-$J$ model.
The model with periodic boundary conditions possesses, 
in addition to the  global $SU(m|n)$
supersymmetry, a hidden dynamical ``quantum group''
symmetry algebra called the 
supersymmetric Yangian \cite{haldane2,yangian,drinfeld}.
This symmetry is responsible
for the ``supermultiplets'' in the energy spectrum 
and the ideal gas-like features of the elementary
excitations and, furthermore, provides us with a simple numerical way
to identify the exact content of the elementary excitations 
relevant for the various correlation functions.

The supersymmetric generalization of the $SU(n)$
Haldane-Shastry model Hamiltonian \cite{lattice,sun1,sun2} is given by
\begin{equation}
H =t \sum_{i < j} {P_{ij}\over d^2(n_i-n_j)},
\label{hamil}
\end{equation}
where $d(x) = (N_a/\pi)\sin(\pi x/N_a)$ and $N_a$ is the total number of sites.
If $a^\dagger_{i\alpha}$ $(a_{i\alpha})$ creates (destroys) a particle of
species $\alpha$ at site $i$ and satisfies the single occupancy condition, 
$\sum_\alpha a^\dagger_{i\alpha}a_{i\alpha} = 1$,
the exchange operator can be written as
$P_{ij}$
= $\sum_{\alpha\beta} a^{\dagger}_{i\alpha}a^{\dagger}_{j\beta}a_{i\beta}
a_{j\alpha}$.
If $m$ of the species labeled by $\alpha$ are bosons, and
$n$ are fermions, the model (\ref{hamil})
has a global $SU(m|n)$ supersymmetry
with generators given by the traceless part of
$J^{\alpha\beta}_0 = \sum_i a^{\dagger}_{i\alpha}a_{i\beta}$.
The Yangian
symmetry generator of the periodic  ISE model is
\begin{equation}
J^{\alpha\beta}_1 = \sum_{i>j,\gamma} w_{ij} a^{\dagger}_{i\alpha}
a^{\dagger}_{j\gamma}a_{i\gamma}a_{j\beta} ,
\end{equation}
where $w_{ij} = \cot(\pi(i-j)/N_a)$.  The higher order generators of 
the Yangian are obtained recursively from various commutators 
involving only $J_0$ and $J_1$ \cite{yangian,drinfeld}.

If we specialize to $SU(1|2)$ supersymmetry, 
with $\alpha  \in \{0, \uparrow , \downarrow \} $, 
we can rewrite the Hamiltonian
in terms of the $SU(2)$ fermionic operators
$c^{\dagger}_{i\sigma} $ = $a^{\dagger}_{i\sigma}a_{i0}$ as
${\cal P} H^0 {\cal P}$, where $H^0$ (up to a shift
in total energy and
in chemical potential) is 
\begin{equation}
-\sum_{i\ne j,\sigma} t_{ij}
c^\dagger_{i\sigma}c_{j\sigma}  + 
\sum_{i<j}\left ( J_{ij} {\bf S}_i \cdot {\bf S}_j  + V_{ij}n_in_j
\right ),
\end{equation}
where $t_{ij}$ = $J_{ij}/2$ = $-2V_{ij}$ = $t/d^2(i-j)$ and
$n_i= n_{i\uparrow}+ n_{i\downarrow}$; 
$\cal P$ is the projection 
operator that projects out all states with doubly-occupied sites.
The ground state $|\Psi_0\rangle $ 
of this model is known \cite{kura,sun1} to be
\begin{equation}
\sum_{\{x,\sigma\}} \prod_{i>j} (z_i-z_j)^
{\delta_{\sigma_i,\sigma_j}} (i)^{{\rm sgn}(\sigma_i-\sigma_j)}
\prod_k z_k^{J_0} \prod_j c^\dagger_{x_j\sigma_i} |0\rangle,
\end{equation}
where $z_j = \exp(i2\pi x_j/N_a)$, $J_0 = -(N/2-1)/2$, $N$ is the
total number of particles, and $|0\rangle$ the electron vacuum
(empty state).
In order to have a non-degenerate ground state,
we take $N/2$ to be odd.
Note that this wave function is just the full Gutzwiller 
projection of a  free electron state\cite{vollhardt}.

A remarkable feature of this model is that the eigenstates
of (\ref{hamil}) form 
degenerate ``supermultiplets''\cite{lattice} 
with multiplicities much higher than those expected from
the global supersymmetry.  
All supermultiplets on the $SU(m|n)$ model
with $m, n > 0 $ 
are present (with the same energy and momentum, but
multiplicity reduced to 2) in the spinless free fermion
$SU(1|1)$ model\cite{haldane2}.
This means that they can be represented by a binary sequence of
$N_a-1$ ones and zeroes, representing (in the spinless fermion model)
the occupations of Bloch states with non-zero momentum (the zero-momentum
orbital has zero energy, which is the supersymmetry, and its
occupation is not fixed).    
There are thus $2^{N_a-1}$ distinct supermultiplets.

In the $SU(1|2)$ case, the 
``occupation number'' sequence describes a supermultiplet
spanning a large range of possible fermion charges $N$.   The state of
minimum charge in the supermultiplet is given by
the number of zeroes in the sequence; the maximum charge is $N_a$ minus the
number of times two consecutive ones occur.  The ground state
of the model with $t > 0$ has a sequence $111 \ldots 111$,
so its minimum charge is $N$ = 0 and its maximum charge is
$N_a - (N_a-2)$ = 2.  The multiplet represented by the alternating 
sequence $10101 \ldots 10101$ has a maximum charge state
$N$ = $N_a$, 
which is the spin-singlet ground state of the antiferromagnetic $S=1/2$
Haldane-Shastry chain, and a minimum charge $(N_a-2)/2$.

We study the model (\ref{hamil}) with $t > 0$ and 
a chemical potential that {\it maximizes} $N$,
so the ground-state has $0 < N < N_a$.
Then, only intermediate states with the {\it maximum}  value of
charge in their supermultiplet contribute to the
thermodynamic limit of the ground-state correlation functions.
To determine the excitation content of these maximal charge states, it
is convenient to add a zero to each end of the binary sequence, 
expanding its length to $N_a + 1$.
The ground state sequence is then
of the form
$0101010 \ldots 1111111 \ldots 0101010$, with a central section
of consecutive ones, with equal-length wings of the alternating
sequence.

In the limit $N = N_a$,
the excitations of the $S=1/2$ antiferromagnet are neutral spin-1/2
{\it spinons}\cite{bethe,fadeev,experm} represented by consecutive zeroes
(e.g. $\ldots 01010010101 \ldots $) and spinless charge $-e$
{\it holons} by consecutive ones (e.g. $\ldots 010101101010 \ldots $).
At intermediate densities, the central
region $\ldots 1111111 \ldots $ may be considered as
a {\it holon condensate} or ``pseudo-Fermi-sea''.   However, the holons
and spinons are {\it not} fermions, but ${\it semions}$, or
particles with
``half-fractional'' statistics, resulting from the spin-charge
separation of a hole, which is a spin-1/2, charge $-e$ fermion.
A configuration $ \ldots 11111110111111 \ldots $ has  a
{\it ``hole in the holon condensate''} which we will call
an ``antiholon'';
because of the semionic statistics of the holons, 
we identify it as  a charge $+2e$, spinless boson.
     
Using concepts from Chern-Simons theory, as
applied to the fractional quantum Hall effect\cite{CS}, 
if condensed particles have charge $q$ and statistics $\Theta$ = $\pi\lambda$, 
vortices or holes
in the condensate have charge $-q/\lambda$ and statistics $\Theta '$ =
$\pi / \lambda $.  Here holons have charge $-e$ and $\Theta$ 
= $\pi /2$ (a semion), so the vortex or hole in the holon condensate 
(antiholon) then has
charge $2e$ and $\Theta = 2\pi $ (a boson).
The applicability of such ``2D'' concepts to
1D ISE-type models has recently been
demonstrated: the holon (antiholon) corresponds to  particle (hole) excitations
of the $\lambda = 1/2$ 
Calogero-Sutherland model where the particle excitations
are semions and the holes $\lambda = 2$ bosons \cite{haldane2,ha}.

The main results of this paper can be summarized in 
Table I, which lists all the possible elementary excitations for
the corresponding local perturbations of the ground state.
The quantum symmetry prevents the injected electron or hole from
breaking up into more than a very simple set of 
elementary excitations consisting of the
left (right) spinons($s_{L(R)}$), holons($h_{L(R)}$), 
and antiholons($\bar h$).  As a result, the
spectral functions of the various dynamical correlation
functions vanish except in certain regions of the 
energy-momentum plane ({\it i.e.}, has {\it ``compact support''}).

Figs.\ \ref{ragreen}-\ref{ss} show the regions of  compact support
formed by the {\it finite} number of elementary excitations 
contributing to the intermediate states for $G^{(1)}$, $C^{(c)}$,
and $C^{(s)}$, respectively.
If the correlation functions are given by the following integral,
\begin{equation}
C(x,t) = \int_{(Q,E)\in \sigma} dQ\ dE\ S(Q,E)\ e^{i(Qx-Et)},
\end{equation}
the figures show the region $\sigma$ where the spectral function
$S(Q,E)$ is non-zero; this is determined by
combining the energies and (Bloch) momenta of the finite
set of elementary excitations contributing to $S(Q,E)$.

The dispersion relations
for the spinon ($E_s$), holon ($E_h$) and antiholon ($E_{\bar h}$)
in the thermodynamic limit are given by 
\begin{mathletters}
\label{dispersion}
\begin{equation}
E_{s_{R(L)}}/t = -q(q \mp v_s^0),\ 0 \le |q| \le {\pi {\bar n}\over 2},
\end{equation}
\begin{equation}
E_{h_{R(L)}}/t = q(q \pm v_c^0), \ 0 \le |q| \le {\pi {\bar n}\over 2},
\end{equation}
\begin{equation}
E_{\bar h}/t = {(v_c^0)^2 - q \over 2},\ -v_c^0 \le q \le v_c^0,
\end{equation}
\end{mathletters}
where $v_s^0 = \pi$ (spin-wave velocity), 
$v_c^0 = \pi(1-{\bar n})$ (sound velocity) and $\bar n$ the density
of electrons.  The right (left) movers take the upper (lower) signs 
and are allowed only in $q \ge 0$ ($q \le 0$) relative 
to the $Q=0$ ground state.  The curvature of the antiholon dispersion
is half that of holon, indicating that {$\bar h$} is made by destroying
two holons.
It is then natural to assign charge $C=+2e$ and $S=0$ to the antiholon while
$C=0$ and $S=\frac{1}{2}$ to the spinon, and $C=-e$ and $S=0$ to the holon.
This assignment is consistent with the results given in Table I and the 
phase shift calculations.
In fact, using this charge conservation argument we were able to 
identify one extra right holon for the local hole excitation 
($\hat{\cal O}_i = c_{i\sigma}$) in Table I,
which could not be resolved numerically because of the small system size
($N_a = 12$) studied.

We outline below how to find the regions
of support for the various correlation functions.
First, we numerically find all the eigenstates 
having non-zero overlap with the states $c_{i\sigma} (c_{i\sigma}^\dagger)
|\Psi_0\rangle$ (for $G^{(1)}$), 
$(n_{i\uparrow}+n_{i\downarrow})|\Psi_0\rangle$ (for $C^{(c)}$)
and $(n_{i\uparrow}-n_{i\downarrow})|\Psi_0\rangle$ (for $C^{(s)}$).
Second, we identify the excitation content of the states by 
inspecting the corresponding motifs where the spinons, holons and
antiholons can easily be identified (see Table I).
We empirically find the following selection rules that the
holon ($v_h$), spinon ($v_s$), antiholon ($v_{\bar h}$), 
spin wave ($v_s^0$) and sound ($v_c^0$) velocities always satisfy:
(i) $v_c^0 < v_s^0$ (i.e. spin-charge separation),
(ii) $v_c^0 \le |v_h| (|v_s|) \le v_s^0$, (iii) $ |v_{\bar h}| \le v_c^0$, 
and (iv) for a given spinon-holon pair $(s_R, h_R)$, 
$|v_{s_R}| \ge |v_{h_R}|$.  
These rules together with the 
results in Table I and Eqs. (\ref{dispersion}) allow us to plot the 
regions of compact support as shown in Figs. \ \ref{ragreen}-\ref{ss}.  

Fig. \ \ref{ragreen} shows the region of support for the 
one-particle Green's function where the states $c_{i\sigma}|\Psi_0\rangle$
($c_{i\sigma}^\dagger|\Psi_0\rangle$) propagate in time 
with positive (negative) energy with respect to the ground state.
The spectral functions should be non-analytic along all the
solid lines where the elementary excitations either ``touch'' the boundaries
or the other elementary excitations.  
When the antiholons are suppressed (i.e. near half filling),
the holon is accompanied either by a spinon
or by three spinons in $S = 1/2$ state.
At $3k_F$ ($2\pi - 3k_F$), where $k_F = \pi {\bar n}/2$, the left (right) 
moving spinon is missing from the
state $c_{i\sigma}^\dagger|\Psi_0\rangle$ since the charge conservation
prevents more than one holon in the presence of one antiholon. Of course,
if two antiholons were  allowed then   states of the type
$(s_L,h_L) + 2{\bar h} + 2(s_R,h_R)$ would  contribute.  
Our numerical study indicates that states
with two antiholons do not contribute.
In fact,  the observed states listed in Table I 
are the simplest possible states satisfying the charge (spin) conservation 
with at most one antiholon.

In Fig.~\ref{dd}, 
only holon-antiholon branches are present at $4k_F$ ($2\pi-4k_F$) 
while the spinon-holon branches show up at $2k_F$ ($2\pi-2k_F$).
At ${\bar n} = 0.1$, the spin-charge separation is hardly visible.
In Fig.~\ref{ss}
we find that the pure spinon excitations are possible only if they both
belong to the same sector, otherwise they are accompanied by
two holons and an antiholon.  The excitation content we find for
$S^z_i(=(n_{i\uparrow}-n_{i\downarrow})/2)$ should be same for $S^{\pm}_i$
since the ground state is a spin singlet. 
As ${\bar n} \rightarrow 0$
we recover the two spinon spectrum for the $S=1/2$ spin chain.

Finally, we have examined how the ISE results for the charge
of the elementary excitations change if we interpolate
between the ISE and NNE $t-J$ models,
which are respectively the $\gamma$ = 0 and $\gamma$ = $\infty$
limits of the integrable family of {\it hyperbolic}
models with exchange $ \propto 1/\sinh^2 \gamma (i-j) $\cite{inozem}.
Away from the ISE limit, the charge carried by the 
holon and antiholon excitations vary with their velocity, and 
become  equal in magnitude (and opposite in sign) as the velocities
approach the sound velocity $v_c^0$.  In the ISE limit, however
the holon charge ($|v| > v_c^0 $) is always $-e$, and the
antiholon charge ($|v| < v_c^0$) is always $+2e$.

The ``dressed charge'' carried by the excitations can be calculated
using the asymptotic Bethe Ansatz\cite{korepin}.
The charge enhancement of the test holon is measured by the 
difference in the phase shifts
of the holon condensate at the pseudo-Fermi points and will in 
general depend on where the holon is with respect to the condensate.  
The total charge C (the bare plus the enhanced)
is plotted in Fig.~\ref{charge} as a function of the 
momentum of the test holon at a fixed density of electrons 
($\bar n = 0.5$) for various values of $\gamma$.  
The pseudo-Fermi points of the condensate for each $\gamma$
are labeled by ``x''.  
The ISE limit is given by the solid line.  
The curve with the smallest charge enhancement in the condensate
corresponds to the NNE model.
In the ISE limit, there is a clear jump in the 
holon charge from $-e$ to $-2e$ 
at the pseudo-Fermi point $\pi(1-\bar n)$.  Therefore, if a holon is 
taken out the condensate, the hole excitation carries charge $+2e$
independent of where it is in the condensate.  We call this hole
an antiholon.  For all the other values of
$\gamma$, there is a considerable charge enhancement of the test
holon in the condensate, and as $\gamma \rightarrow 0$ the 
charge approaches $-2e$.

In conclusion, we have devised simple rules for
constructing the motifs for the excited states 
of the 1D ISE $t$-$J$ model
and identified the exact excitation content
of the intermediate states for the one-particle Green's function,
the charge density-density and spin-spin correlation functions.
We believe that this model is in the same universality class
as the NNE model, and that the most relevant states for the ground state
correlation functions of the NNE model are also given by Table I.
Finally, the presence of spinons, holons, and antiholons in two-dimensional
models and their role in the high $T_c$ superconductivity 
is an amusing possibility.

ZNCH is supported by DOE grant \#DE-FG02-90ER40542 and FDMH by
NSF-DMR-922407.

\begin{figure}
\caption{{\it Compact support} for the one-particle Green's
function.  The momentum is in units of $\pi$ and the excitation energy E
in $\pi^2/t$. The contributing elementary excitations to
this region are $(h_L,s_L) + {\bar h} + 2(h_R,s_R)$ 
for the positive energy part 
(i.e. $c_{i\sigma}|\Psi_0\rangle$) and $(s_L,h_L) + {\bar h}$ for the 
negative part (i.e. $c_{i\sigma}^\dagger|\Psi_0\rangle$).  
Their mirror states (i.e. $L$ and $R$ exchanged) also contribute.
The four momenta at which
$E=0$ is allowed are $k_F$, $2\pi-3k_F$,
$3k_F$, and $2\pi-k_F$ where $k_F = \pi {\bar n}/2$.}
\label{ragreen}
\end{figure}

\begin{figure}
\caption{{\it Compact support} for the density-density correlation
function.  $(s_L,h_L) + {\bar h} + (s_R,h_R)$,
${\bar h} + 2h_R$ and their mirror states contribute. $E=0$ is 
allowed at $0$($2\pi$), $2k_F$, $2\pi-4k_F$, $4k_F$, $2\pi-2k_F$.
Only holon-antiholon branches are present at $4k_F$ ($2\pi-4k_F$) indicating
that $4k_F$ is the holon Fermi point.}
\label{dd}
\end{figure}

\begin{figure}
\caption{{\it Compact support} for the spin-spin correlation function.
$(s_L,h_L) + {\bar h} + (s_R,h_R)$, $2s_L$ and their mirror states
contribute. 
$E=0$ allowed at $0(2\pi)$, $2k_F$, $2\pi-2k_F$. This indicates
that $2k_F$ is the spinon Fermi point.}
\label{ss}
\end{figure}

\begin{figure}
\caption{Charge of a test holon versus its momentum 
in the vicinity of the holon condensate 
for $\gamma = $ 0, 0.2, 0.3, 0.5, 1.0, 2.0, $\infty$.  The charge is
in units of $-e$ where $e$ is the electron charge and the momentum of the
test holon in units of $\pi$.
The pseudo-Fermi points are labeled by ``x'' for each $\gamma$.
The step function corresponds to the ISE model ($\gamma = 0$).  The NNE model
has the smallest but still considerable charge enhancement in
the condensate ($\gamma = \infty$).}
\label{charge}
\end{figure}

\narrowtext
\begin{table}
\caption{List of all the possible excitations from the ground
state perturbed by the local operators 
$c_{i\sigma}(c_{i\sigma}^\dagger)$ ($G^{(1)}$), 
$n_{i\uparrow} + n_{i\downarrow}$ ($C^{(c)}$),
and $n_{i\uparrow}-n_{i\downarrow}$ ($C^{(s)}$). 
The mirror states ($L \leftrightarrow R$) not listed are also allowed.  
The spinon ($v_s$), holon ($v_h$), 
antiholon ($v_{\bar h}$), spin-wave ($v_s^0$) and sound ($v_c^0$) 
velocities always satisfy: 
(i) $v_c^0 < v_s^0$,
(ii) $v_c^0 \le |v_h| (|v_s|) \le v_s^0$, (iii) $ |v_{\bar h}| \le v_c^0$, 
and (iv) for a given spinon-holon pair $(s_R, h_R)$, 
$|v_{s_R}| \ge |v_{h_R}|$.   }

\begin{tabular}{dd}
Local Operator $\hat {\cal O}_i$ & Excitation contents of 
$\hat {\cal O}_i|\Psi_0\rangle$ \\
\tableline
$c_{i\sigma}$ & $(s_L,h_L) + {\bar h} + 2(s_R,h_R)$ \\
\tableline
$c^\dagger_{i\sigma}$ &$(s_L,h_L) + {\bar h}$ \\
\tableline
$n_{i\uparrow}+n_{i\downarrow}$ & $(s_L,h_L) + {\bar h} + (s_R,h_R)$\\
& ${\bar h} + 2h_R$\\
\tableline
$n_{i\uparrow}-n_{i\downarrow}$ & $(s_L,h_L) + {\bar h} + (s_R,h_R)$\\
& $2s_L$\\
\end{tabular}
\label{tab1}
\end{table}
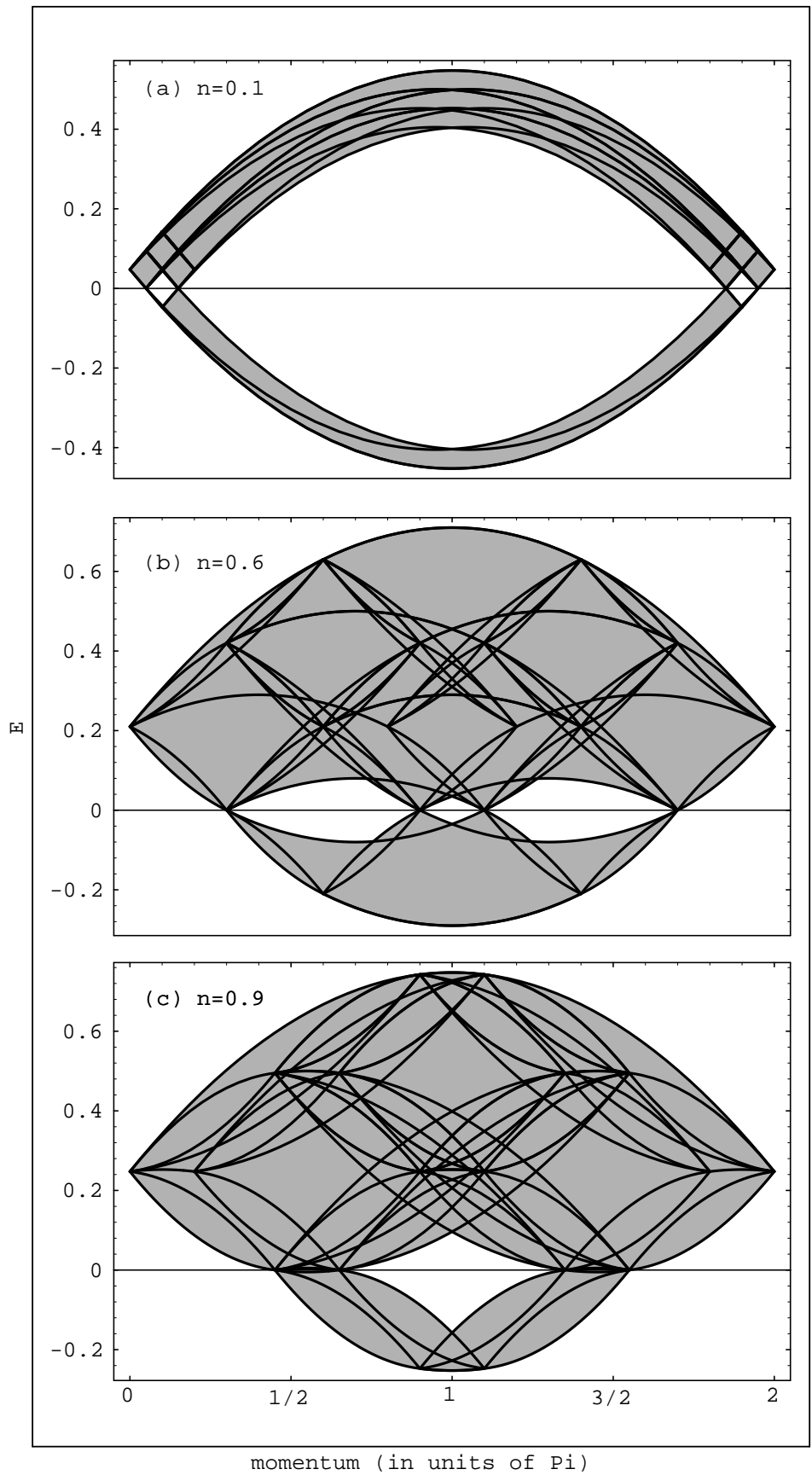
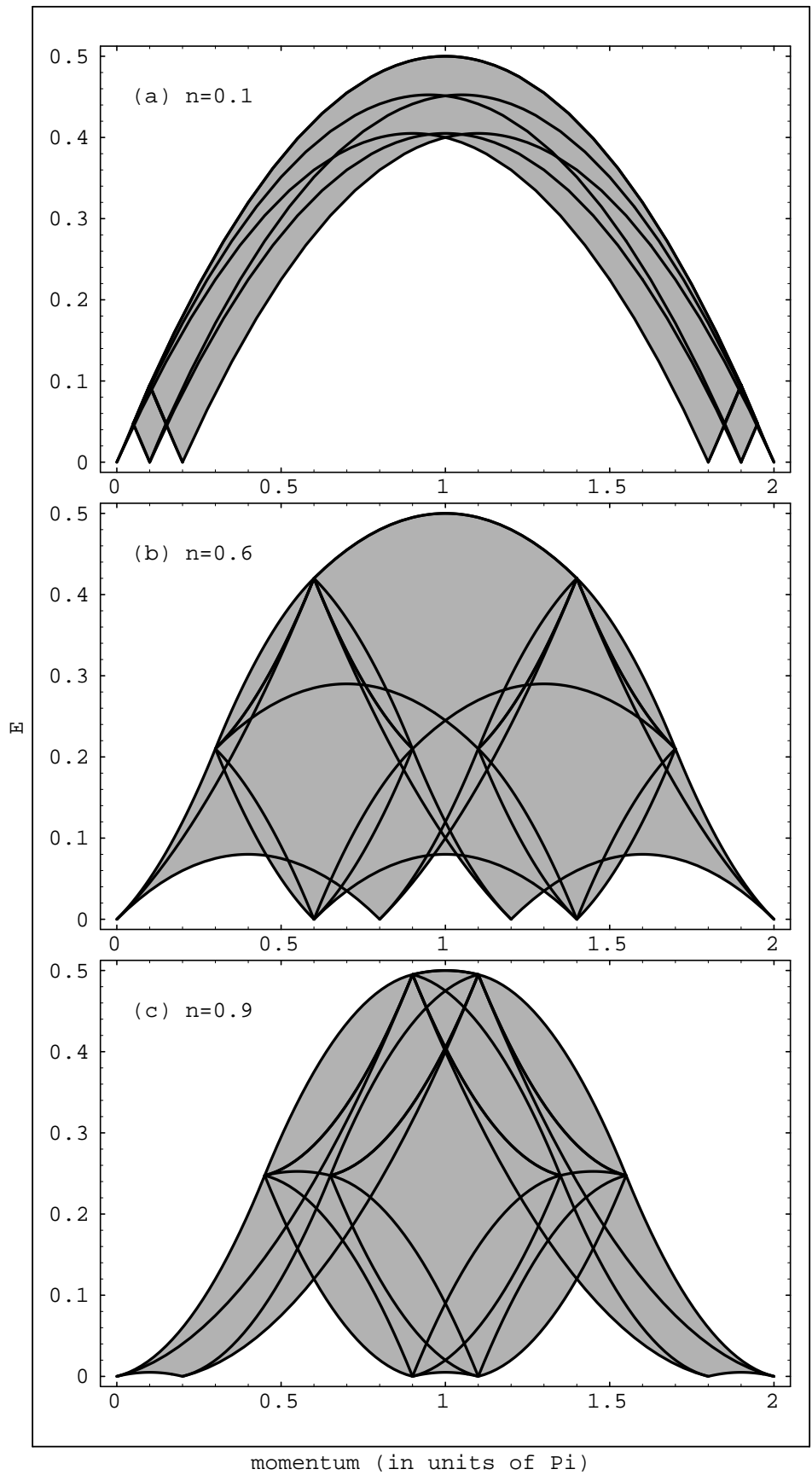
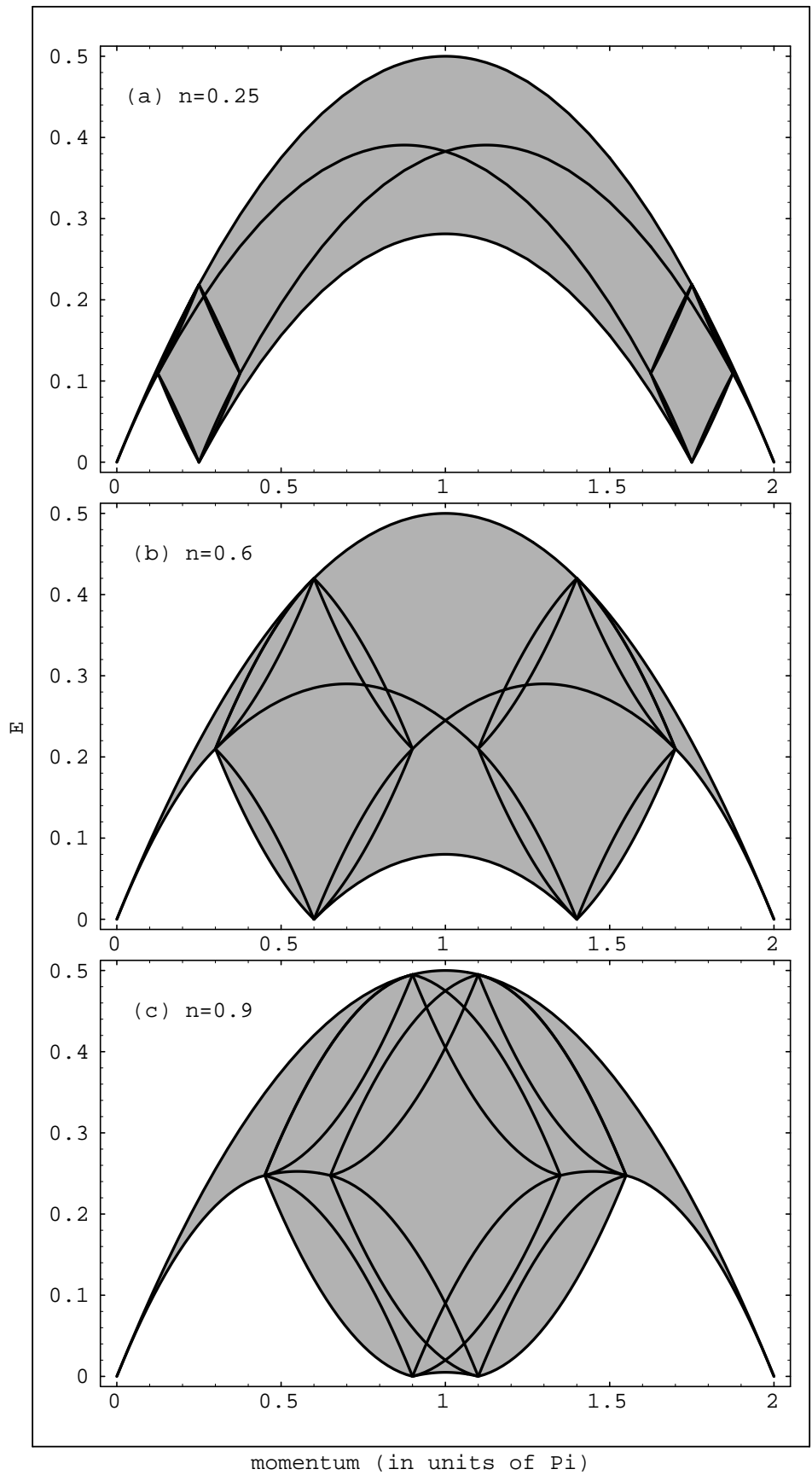
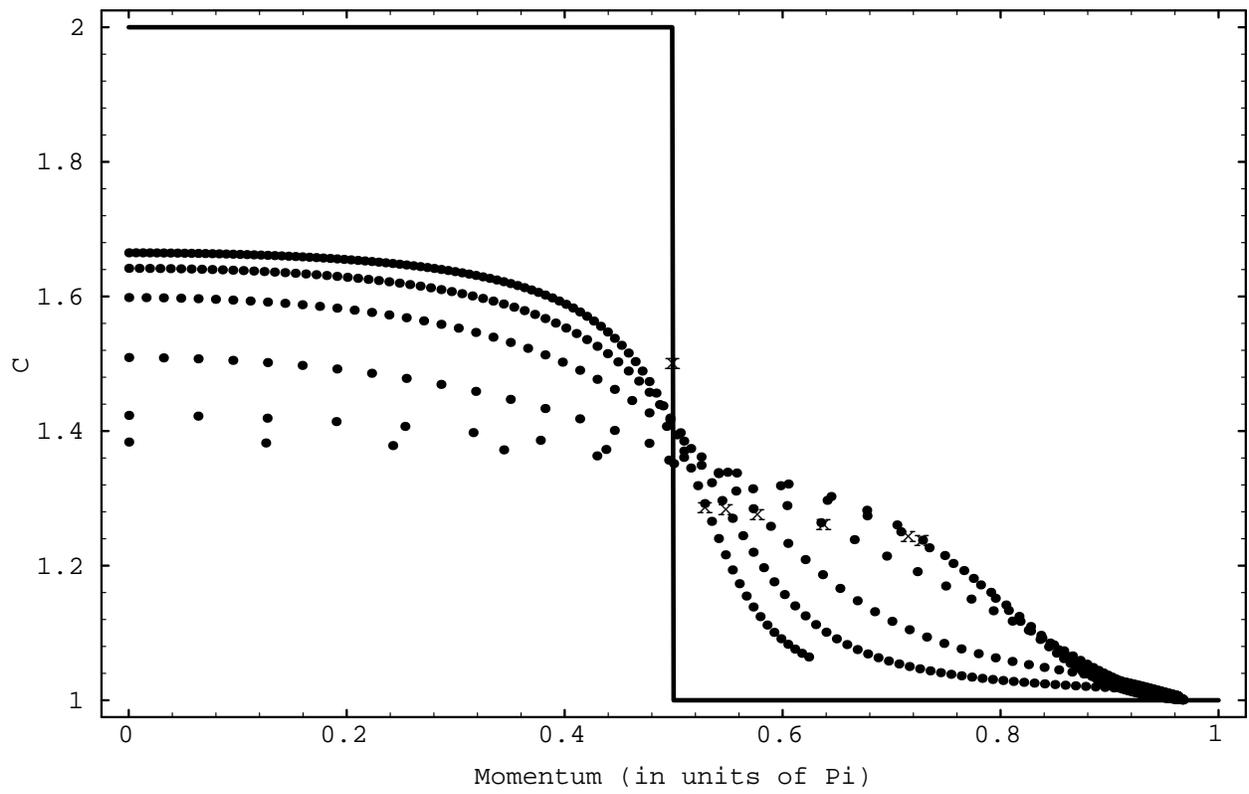
\end{document}